# 毫米波低功耗混合波束赋形架构及算法

张洪浦[1]，郭玉路[1]，薛刘荀[1]，刘星晨[1]，孙舒[1,**]，高锐锋[2,3]，喻翔昊[4]，陶梅霞[1]
（[1]上海交通大学电子工程系与未来媒体网络协同创新中心，上海 200240；
[2]南通大学交通学院，南通 226019；
[3]南通先进通信技术研究院有限公司，南通 226019
[4]香港城市大学电机工程学系，香港 999077）

【摘要】本文研究毫米波通信系统中的低功耗混合波束赋形架构及其算法设计，旨在应对现有毫米波混合波束赋形中硬件灵活性低与功耗高的挑战。为解决现有混合波束赋形架构的不足，本文提出一种新型低功耗混合波束赋形架构，该架构通过在移相器网络前后引入射频开关网络，实现射频链路与移相器及天线阵列的动态连接，并建立所提架构的系统模型，包括数字预编码与模拟预编码过程，同时考虑数字模拟转换器的量化误差与移相器分辨率等硬件限制。为最大化能源效率，本文推导了包含频谱效率和系统功耗的能源效率模型，提出基于块坐标下降的混合预编码算法，迭代优化数字预编码、模拟预编码矩阵及数模转换器分辨率。基于 NYUSIM 毫米波信道仿真平台的仿真结果表明，所提混合波束赋形架构与预编码算法在完整与部分信道状态信息下的能源效率均优于现有的代表性架构及预编码算法，同时频谱效率相较于全连接架构的损失低于 20%。

# MILLIMETER-WAVE ENERGY-EFFICIENT HYBRID BEAMFORMING ARCHITECTURE AND ALGORITHM

ZHANG Hongpu[1], GUO Yulu[1], XUE Liuxun[1], LIU Xingchen[1],
SUN Shu[1,**], GAO Ruifeng[2,3], YU Xianghao[4], TAO Meixia[1]
（[1]Department of Electronic Engineering and the Cooperative Medianet Innovation Center, Shanghai Jiao Tong University, Shanghai 200240, China;
[2]School of Transportation, Nantong University, Nantong 226019, China;
[3]Nantong Research Institute for Advanced Communication Technologies, Nantong 226019, China
[4]Department of Electrical Engineering, City University of Hong Kong, Hong Kong 999077, China）

【Abstract】This paper studies energy-efficient hybrid beamforming architectures and its algorithm design in millimeter-wave communication systems, aiming to address the challenges faced by existing hybrid beamforming due to low hardware flexibility and high power consumption. To solve the problems of existing hybrid beamforming, a new energy efficient hybrid beamforming architecture is proposed which introduces radio-frequency (RF) switch networks at the front and rear ends of the phase shifter network, enabling dynamic connections between the RF chains and the phase shifter array as well as the antenna array. The system model of the proposed architecture is established, including digital precoding and analog precoding processes, considering practical hardware limitations such as quantization errors of the digital-to-analog converter (DAC) and minimum resolution of phase shifters. In order to maximize the energy efficiency of the transmitter, this paper derives an energy efficiency model that includes spectral efficiency and system power consumption, and proposes a hybrid precoding algorithm based on block coordinate descent. The algorithm iteratively optimizes the digital precoding matrix, analog precoding matrix, and DAC resolution to improve the overall energy efficiency of the system. Simulation results under the NYUSIM-generated millimeter-wave channels show that the proposed architecture and algorithm have higher energy efficiency than existing representative architectures and precoding algorithms under complete and partial channel state information, while the loss of spectral efficiency compared to fully connected architectures is less than 20%.
【Key words】Millimeter-wave, hybrid beamforming, low power consumption, block coordinate descent, alternating minimization

*基金项目：
国家自然科学基金“毫米波全息超表面的无线信道特性研究”（62271310）
国家自然科学基金“低空信道智能建模与预测推理”（62431014）
中央高校基本科研业务费专项资金
南通市自然科学基金“面向海洋通信网络的平滑低时延分组传输方法”（JC2023074）
香港研究资助局优配研究金（16212922）
卫星导航技术重点实验室（WXDHS2023105）
**通信作者

# 1　引言

随着近年来无线通信设备数量的快速增长和用户需求的多样化，现有的无线通信技术面临着日益复杂的挑战。为了满足未来通信系统对于超高吞吐量、更大带宽、更低时延和更高可靠性的需求，下一代无线通信亟需开发新的频段资源。目前主流的无线通信技术大多使用 6GHz 以下的频段，频谱资源已十分拥挤；而毫米波频段（30GHz - 300GHz）拥有远超 sub-6GHz 频段的丰富频谱资源与可用带宽，为实现大带宽、低时延、高速率的通信提供了较为理想的条件，被视为缓解频谱资源短缺的有效解决手段。同时，毫米波频段的低时延优点可赋能多方面的技术应用[1] 包括局域网与个人区域网络、车辆协作通信与感知等，在 5G 乃至未来 6G 网络中具有广阔的应用前景[2-4]。

毫米波通信也面临诸多技术挑战：与 sub-6GHz 频段相比，毫米波波长较短，其在自由空间传输中的损耗相对较高，衍射能力较弱；某些波段易受降水和大气衰减的影响[1]，这些因素都会对通信质量造成一定程度的影响。毫米波频段的高频与宽带特性亦对硬件电路设计提出了严格的要求，例如高功耗和高成本的射频组件。毫米波频段通信的关键特征之一是结合大规模多输入多输出（Massive Multiple Input Multiple Output, Massive MIMO）技术，在发射端与接收端上使用大规模天线阵列[5]，这些阵列用于提供足够的阵列增益，合成高指向性的波束，能够有效提高通信系统的频谱效率（Spectral Efficiency, SE）并合理利用带宽[1,6,7]。在通信系统中，波束赋形（预编码）技术通常与 Massive MIMO 同时使用来提供定向的波束与相关增益。文献 [8] 研究了毫米波通信系统中的 Massive MIMO 部署，有效验证了波

束赋形与 Massive MIMO 在 5G 网络中的应用。目前大量研究集中在不同的波束赋形架构以及预编码算法设计上，以提高毫米波 Massive MIMO 的频谱效率与能量效率。

目前，波束赋形技术可分为全数字波束赋形、模拟波束赋形和数字模拟混合波束赋形三类。在 sub-6GHz 频段的 MIMO 系统中使用的往往是全数字波束赋形技术，从本质上说全数字波束赋形是数字信号处理的实际应用[9]，其对信号的赋形权值与相移完全由数字电路实现，因此需要在每个天线阵列元前都配置一条射频链路，每条射频链路都包含混频器、数模转换器/模数转换器（Digital to Analog Converter/Analog to Digital Converter, DAC/ADC）以及功率放大器等组件[10]，在毫米波中实现该方案需要数量庞大的硬件器件，会带来显著的成本与功耗。模拟波束赋形完全依靠射频域处理以此减少射频链数量[11]，它在模拟域使用模拟移相器网络来改变馈送到不同天线阵列元的信号的相对相位，从而在指定的方向上发射或接收对应的波束，其中模拟域由一条射频链路与若干模拟移相器（Phase Shifter, PS）组成，但由于其只包含一条射频链路，在毫米波通信中只适用于单用户和单数据流的传输，并不能充分利用可用的空间频谱资源。为此，混合波束赋形技术应运而生，它结合了全数字和模拟波束赋形的优势，既能减少射频链路数量，降低硬件复杂度，又能在多用户、多流通信场景中保持较高的频谱效率[12-13]。其通过把波束赋形过程分为数字域与模拟域两个阶段来实现，并通过有限数量的射频链路来连接数字预编码器与模拟预编码器。

从混合波束赋形架构的硬件连接结构上来看，各条基带上的射频链路可被设计为与所有的天线阵列元相连、与一定分组的天线阵列元相连或是与通过一定优化算法划分出的天线阵列子集相连；三种不同的连接结构分别被称为全连接架构[14-15]、部分连接架构[16]与动态子阵列连接架构[17-18]，针对这些架构又衍生出不同的预编码算法。全连接架构虽能较好地逼近全数字预编码架构下的频谱效率，且在硬件复杂度有一定程度上的降低，但由于该架构下所用模拟移相器的数量与天线阵列规模与射频链路数量成正比，导致系统功耗仍较高，系统能源效率较低。部分连接架构中的射频链路只需与天线阵列的一个固定子集相连接，与全连接架构相比，该架构具有硬件复杂度较低、易于实现等优点，但同时由于射频链路的潜能未能完全利用，会造成一定的性能损失。动态子阵列混合波束赋形架构是目前研究的一个热点架构[17-18]。该架构在模拟波束赋形部分开创性地采用低功耗以及低成本的射频开关网络与移相器网络相结合的方法。在这种架构下，每个天线阵列元前连接一个自适应调节移相器，通过开关网络选择一个射频链路相连，此架构中使用的移相器数量等于天线元的数量，远小于全连接架构。相较于部分连接架构，该架构中射频链路与天线阵列子集的连接有更多的自由度，从而在不损失过多频谱效率的情况下有更低的功耗与较高的能源效率。

虽然部分连接架构与动态子阵列架构的系统功耗相较于全连接架构已有不少的优化，但目前混合波束赋形架构及优化算法仍存在一些问题尚待研究。首先，架构灵活性仍有不足。动态子阵列架构虽降低了系统中移相器的数量，但每个天线阵列元只能通过一个移相器与一个射频链路相连的映射结构仍有些固定，在 Massive MIMO 系统中的功耗较大。其次，目前研究大多聚焦与混合波束赋形硬件上的单个问题，鲜有研究共同考虑带有量化相位的移相器和 DA/ADC 分辨率对各种混合波束赋形架构频谱效率的影响。另外，不同 CSI 情况下的各混合波束赋形架构的预编码算法不完全，目前基于部分 CSI 下的预编码算法设计主要针对全连接架构与部分连接架构，对于更为灵活的混合波束赋形架构来说需要设计相应的预编码算法。

针对上述问题，本文设计一种基于射频开关与移相器组合的混合波束赋形架构。同时，对于提出的混合波束赋形连接架构，结合较为全面和实际的模拟硬件限制，推导窄带单用户场景下的最大化能源效率问题，并基于完整的信道状态信息设计预编码算法，使得系统在功耗与频谱效率中达到一定的平衡。通过在 NYUSIM 信道下的仿真结果的分析，证明了所提出结构的低功耗与算法的有效性。

符号标记：小写和大写粗体字母分别表示向量和矩阵；$(\cdot)^T$ 和$(\cdot)^H$ 分别表示矩阵的转置与共轭转置；$\|\cdot\|_0$ 和$\|\cdot\|_F$ 分别表示矩阵的 0 范数和 Frobenius 范数；Re$\{\cdot\}$和 Im$\{\cdot\}$分别表示复数的实部和虚部；Arg$(\cdot)$表示复数的主辐角；$\mathbf{I}_N$ 表示 $N\times N$ 的单位矩阵；Tr$(\cdot)$表示矩阵的迹；$\mathbb{E}(\cdot)$表示数学期望；$\mathcal{B}$表示元素取值为 0 或 1 的矩阵的集合空间；$\mathbb{C}$表示元素取值为复数的矩阵的集合空间。

# 2　系统模型

## 2.1　基于开关与移相器的混合波束赋形架构模型

针对目前混合波束赋形架构中硬件连接趋于固定的问题，本文提出一种新型的混合波束赋形架构，该架构在移相器网络的前后两端均引入射频开关网络，从而实现更加灵活的硬件连接。在发射端的此种混合波束赋形架构如图 1 所示

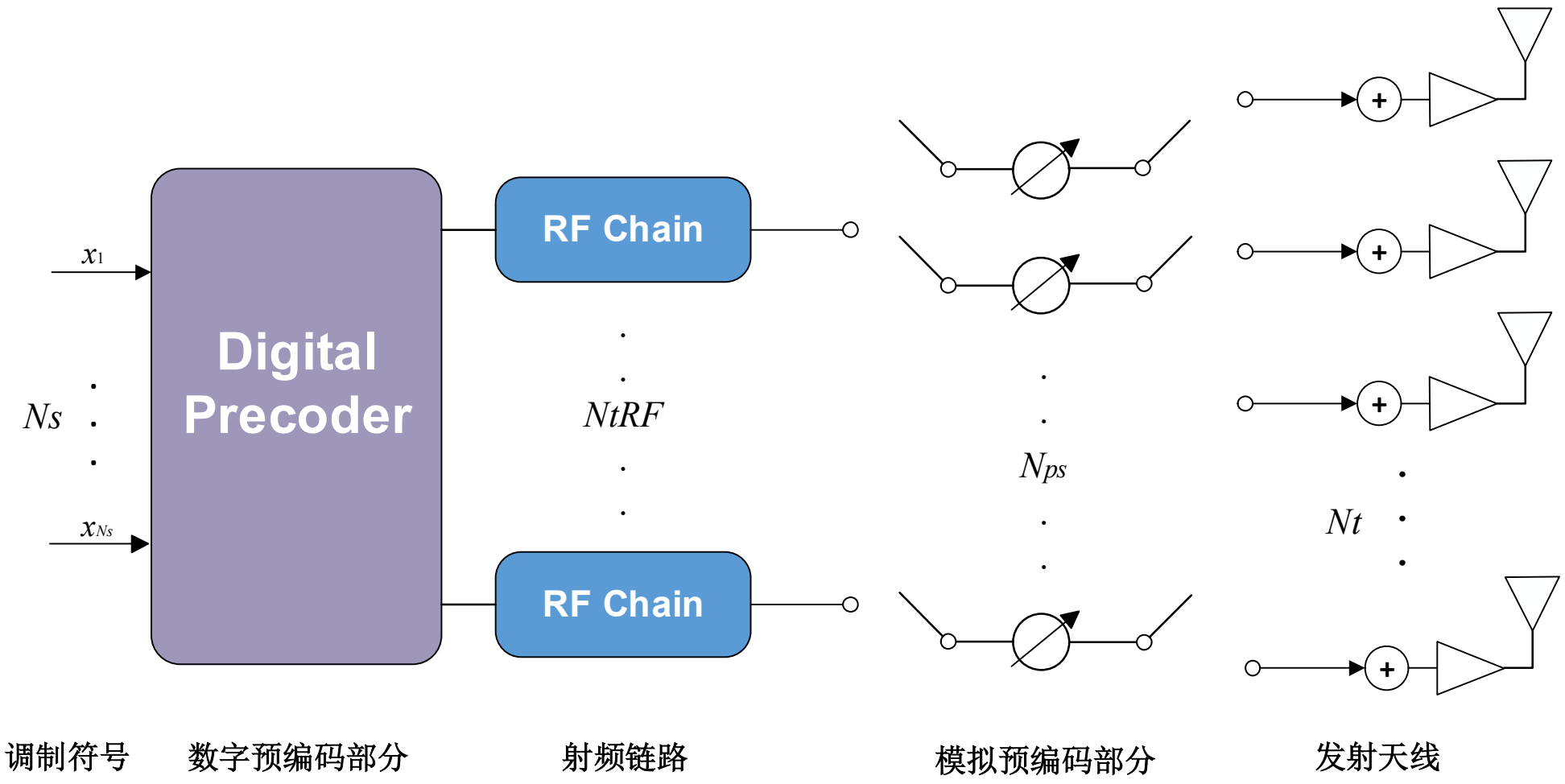

图 1　　发射端带有两端开关网络的混合波束赋形架构图

考虑使用图 1 中描述的混合波束赋形架构的单用户毫米波 MIMO 通信系统的下行传输场景，其中发射端配备有 $N_t$ 根天线，发射端向接收端发送 $N_s$ 个数据流。在发射端同时配备有 $N_{tRF}$ 条射频链路。在上述通信系统中，设需要传输的数据流符号向量是维度为 $N_s\times 1$ 的 $\mathbf{x}$，本文假设其满足广泛使用的归一化信号功率，即：

$$\mathbb{E}\left(\mathbf{x}\mathbf{x}^H\right)=\frac{1}{N_s}\mathbf{I}_{Ns} \tag{1}$$

传输信号先经过数字预编码（基带编码）$\mathbf{F}_{\mathrm{BB}}\in\mathbb{C}^{NtRF\times Ns}$ 进行数字信号处理，之后经过模拟预编码 $\mathbf{F}_{\mathrm{RF}}\in\mathbb{C}^{Nt\times NtRF}$ 进行相位调整。在该混合波束赋形架构中，模拟预编码矩阵 $\mathbf{F}_{\mathrm{RF}}$ 可以表示为：

$$\mathbf{F}_{\mathrm{RF}}=\mathbf{S}_2\mathbf{F}_{\mathrm{ps}}\mathbf{S}_1 \tag{2}$$

其中 $\mathbf{S}_1\in\mathcal{B}^{Nps\times NtRF}$ 为负责移相器与射频链路连接的前置开关矩阵，$\mathbf{F}_{\mathrm{ps}}\in\mathbb{C}^{Nps\times Nps}$ 为提供相移

的可调移相器矩阵，$\mathbf{S}_1 \in \mathcal{B}^{Nt\times Nps}$为负责移相器与天线阵列元连接的后置开关矩阵。在发射端还有总功率上的约束，即：

$$\left\|\mathbf{F}_{\mathrm{RF}}\mathbf{F}_{\mathrm{BB}}\right\|_F^2 = N_s \tag{3}$$

实际应用中的 DAC 器件有一定的分辨率限制，并引入量化误差，且较高分辨率的 DAC 会带来较高的功耗[19]。本文考虑使用线性加性量化噪声模型[20]来表示量化误差，则经过数字预编码后在 DAC 输出端口的信号可表示为：

$$Q\left(\mathbf{F}_{\mathrm{BB}}\mathbf{x}\right) = \mathbf{\Delta}\mathbf{F}_{\mathrm{BB}}\mathbf{x} + \boldsymbol{\varepsilon}_b \tag{4}$$

$\mathbf{\Delta}$ 表示每个射频链路上的 DAC 由于量化误差导致的乘法失真参数，对角元素可表示为：

$$\mathbf{\Delta}_{i,i} = 1 - \frac{\pi\sqrt{3}}{2}2^{-2b_i} \tag{5}$$

$b_i$ 为第 $i$ 个射频链路中 DAC 的分辨率，这里假设所有 DAC 的分辨率范围相同：

$$b_i \in \left[b_{min}, b_{max}\right],\quad i = 1,2,\ldots,N_{tRF} \tag{6}$$

$\boldsymbol{\varepsilon}_b$ 为加性量化噪声，服从复高斯分布，值得注意的是，目前波束赋形技术中主流的 DAC 分辨率约为 8 比特[21]，此时 $\boldsymbol{\varepsilon}_b$ 的协方差矩阵的非零元素的数量级在 $10^{-5}$ 左右，可将其忽略。

在模拟预编码部分大量使用的移相器实际上也并无较高的分辨率，假设在混合波束赋形系统中所使用的移相器拥有量化分辨率 $q$，令 $\Phi$ 表示其能提供的相移，则有：

$$\Phi \in \left\{0, \frac{2\pi}{2^q}, \ldots, \frac{2\pi\left(2^q - 1\right)}{2^q}\right\} \tag{7}$$

在所提出的架构下，$\mathbf{F}_{\mathrm{ps}}$ 为一个对角矩阵，对角元素可表示为：

$$\mathbf{F}_{\mathrm{ps}i,i} = e^{j\Phi_i},\quad i = 1,2,\ldots,N_{ps} \tag{8}$$

考虑上述实际硬件限制后，设在接收端接收到的信号为 $\mathbf{y}$，此处假设信道状态信息（Channel State Information, CSI）是已知并且完整的，则 $\mathbf{y}$ 可以表示为：

$$\mathbf{y} = \sqrt{\rho}\mathbf{H}\mathbf{S}_2\mathbf{F}_{\mathrm{ps}}\mathbf{S}_1\mathbf{\Delta}\mathbf{F}_{\mathrm{BB}}\mathbf{x} + \mathbf{n} \tag{10}$$

其中 $\mathbf{H}$ 是维度为 $N_r\times N_t$ 的信道状态信息矩阵，$\rho$ 为平均接收功率，$\mathbf{n}$ 为毫米波信道夹杂的加性高斯白噪声向量，每个元素独立且服从相同的复高斯分布$\mathcal{CN}(0, \sigma^2{}_n)$，设接收端的解码矩阵为 $\mathbf{W}\in\mathbb{C}^{Nr\times Ns}$，则接收信号可表示为：

$$\hat{\mathbf{y}} = \mathbf{W}^H\mathbf{y} = \sqrt{\rho}\mathbf{W}^H\mathbf{H}\mathbf{S}_2\mathbf{F}_{\mathrm{ps}}\mathbf{S}_1\mathbf{\Delta}\mathbf{F}_{\mathrm{BB}}\mathbf{x} + \mathbf{W}^H\mathbf{n} \tag{11}$$

此时系统的频谱效率可表示为[22]

$$R = \log_2\left|\mathbf{I}_{N_s} + \frac{\rho}{N_s}\left(\sigma_n^2\mathbf{W}^H\mathbf{W}\right)^{-1}\mathbf{W}^H\mathbf{H}\mathbf{S}_2\mathbf{F}_{\mathrm{ps}}\mathbf{S}_1\mathbf{\Delta}\mathbf{F}_{\mathrm{BB}}\mathbf{F}_{\mathrm{BB}}^H\mathbf{\Delta}^H\mathbf{S}_1{}^H\mathbf{F}_{\mathrm{ps}}{}^H\mathbf{S}_2{}^H\mathbf{H}^H\mathbf{W}\right| \tag{12}$$

### 2.2 毫米波信道模型

本文采用能较为准确地描述毫米波信道的扩展Saleh-Valenzu信道模型[23-24]，具体而言，毫米波 CSI 矩阵 $\mathbf{H}\in\mathbb{C}^{Nr\times Nt}$ 可描述为：

$$\mathbf{H} = \gamma\sum_{i=1}^{L}\alpha_i\mathbf{a}_{ri}\mathbf{a}_{ti}^H = \gamma\mathbf{A}_r\mathbf{\Lambda}\mathbf{A}_t^H \tag{13}$$

其中 $\gamma$ 为归一因子，可表示为 $\sqrt{N_tN_r/L}$，$L$ 表示信道中多径的数量，$\alpha_i$ 表示第 i 条多径的复路径增益，$\mathbf{\Lambda}\in\mathbb{C}^{L\times L}$ 是描述多径复增益的对角矩阵，第 $i$ 个对角元素为 $\alpha_i$。$\mathbf{a}_{ri}$ 和 $\mathbf{a}_{ti}$ 分别表示第 $i$ 条多径下的接收端与发射端的阵列响应向量，分别是矩阵 $\mathbf{A}_r$ 和 $\mathbf{A}_t$ 的第 $i$ 列。这里的阵列响应向量仅仅为与天线阵列结构相关的函数，不失一般性，本文考虑使用均匀平面阵列（Uniform Planar Array, UPA）来作为混合波束赋形架构中天线阵列的结构，那么在发射端的阵列响应向量可以描述为

$$\mathbf{a}_{ti}(\phi,\theta) = \frac{1}{\sqrt{N_t}}[1,\ldots,e^{jkd(m\sin\phi\sin\theta+n\cos\theta)},\ldots,e^{jkd\left(\left(\sqrt{N_t}-1\right)\sin\phi\sin\theta+\left(\sqrt{N_t}-1\right)\cos\theta\right)}]^T \tag{14}$$

其中 $k = 2\pi/\lambda$，$\lambda$ 为毫米波的波长，$d$ 为相邻的天线阵列元之间的间距，设置为 $\lambda/2$，$\phi$ 和 $\theta$ 分别为某条多径相较于发射天线阵列平面的出发方位角与仰角方向。$0 \le m < \sqrt{N_t}$ 和 $0 \le n < \sqrt{N_t}$ 分别为天线阵列在 y 轴与 z 轴上的阵列元索引，发射天线总数为 $N_t$。由于接收端的天线阵列同样采用 UPA 架构，故阵列响应向量与式（14）类似，此时天线总数为 $N_r$。

### 2.3 系统功耗模型

一般情况下，发射端混合波束赋形系统的总功耗由数字信号处理硬件功耗、模拟预编码中的射频硬件功耗和射频信号辐射功耗组成[25]。对于数字信号处理硬件组件，该部分的功耗主要是由基带信号处理而产生的。这里可假设数字信号处理的功耗不变，用 $P_{\mathrm{BB}}$ 来表示。

对于模拟预编码部分的射频硬件功耗，如图 1 所示，信号经过基带处理后，先通过射频链路将基带信号转换为射频模拟信号。之后经过模拟预编码部分进行相位调整。模拟预编码部分主要由前置开关阵列，移相器阵列与后置开关阵列组成，前置开关可以选择射频链路的输出端口，其状态决定任意射频链路的输出信号是否通过与其相接的移相器。后置开关矩阵选择天线阵列元进行连接，将经过移相器进行相移后的射频信号馈送到选择的天线，经过天线前的功率放大器（Power Amplifier，PA）后发送到信道中。在实际中，除了 DAC 的功耗会随其所使用的分辨率而改变，其余的射频硬件组件可以被假设为功耗固定。

上述硬件中，发射天线前的 PA 的功耗可以用发射功率的线性形式来表示：

$$P_{\mathrm{PA}} = \frac{P_{\mathrm{t}}}{\rho_{pa}} \tag{15}$$

$P_{\mathrm{t}}$ 代表天线阵列元的发射功率，$\rho_{pa}$ 表示功率放大器的转换效率。

DAC 的功耗是采样频率 $f_s$ 和转换器品质因数 FoM 的线性函数，并随着其分辨率 $b$ 的增加呈指数增长[26]，可表示为：

$$P_{\mathrm{DAC}} = \mathrm{FoM}\times f_s\times 2^b = P_{\mathrm{D}}\times 2^b \tag{16}$$

在该架构中，射频链路和天线之间的连接是由两处的开关阵列决定的，当某条通路被激活时，该通路上的射频链路以及天线会消耗其额定功率，反言之，若某条射频链路或天线上无信号通路，则可将其关闭以节省功耗。故在模拟预编码部分的总功耗可以表示为：

$$P_{\mathrm{RF}} = P_{\mathrm{PS}} N_{ps} + 2P_{\mathrm{SW}} N_{ps} + P_{\mathrm{RFC}} \left\| \mathbf{c}\left(\mathbf{S}_1\right) \right\|_0 + P_{\mathrm{D}} \sum_{i \in \mathcal{C}} 2^{b_i} + \frac{1}{\rho_{pa}} P_t \left\| \mathbf{r}\left(\mathbf{S}_2\right) \right\|_0 \tag{17}$$

其中 $\mathbf{c}(\mathbf{S}_1) = [c_1, c_2, \ldots, c_{NtRF}]$，$\mathbf{r}(\mathbf{S}_2) = [r_1, r_2, \ldots r_{Nt}]^T$，且有：

$$c_n = \sum_{i=1}^{N_{ps}} \mathrm{S}_{1i,n} \tag{18}$$

$$r_n = \sum_{i=1}^{N_{ps}} \mathrm{S}_{2n,i} \tag{19}$$

对于辐射功耗，其主要由传输功率组成，实际上的发送功率主要由发射端的天线提供，可认为单个天线阵列元能提供的功率是平稳的，在所提出的架构中，并非所有天线都被连接，故辐射功耗可表示为：

$$P_{\mathrm{Tr}} = \left\| \mathbf{F}_{\mathrm{RF}} \mathbf{F}_{\mathrm{BB}} \right\|_F^2 = N_s \frac{\left\| \mathbf{r}\left(\mathbf{S}_2\right) \right\|_0}{N_t} \tag{20}$$

系统总功耗可表示为：

$$\begin{aligned} P &= P_{\mathrm{BB}} + P_{\mathrm{RF}} + P_{\mathrm{Tr}} \\ &= P_{\mathrm{BB}} + P_{\mathrm{PS}} N_{ps} + 2P_{\mathrm{SW}} N_{ps} + P_{\mathrm{RFC}} \left\| \mathbf{c}\left(\mathbf{S}_1\right) \right\|_0 \\ &\quad + P_{\mathrm{D}} \sum_{i \in \mathcal{C}} 2^{b_i} + \frac{1}{\rho_{pa}} P_t \left\| \mathbf{r}\left(\mathbf{S}_2\right) \right\|_0 + N_s \frac{\left\| \mathbf{r}\left(\mathbf{S}_2\right) \right\|_0}{N_t} \\ &= P_{\mathrm{con}} + P_1\left(\mathbf{S}_1, \boldsymbol{b}\right) + P_2\left(\mathbf{S}_2\right) \end{aligned} \tag{21}$$

其中

$$P_{\mathrm{con}} = P_{\mathrm{BB}} + P_{\mathrm{PS}} N_{ps} + 2P_{\mathrm{SW}} N_{ps} \tag{22}$$

$$P_1\left(\mathbf{S}_1, \boldsymbol{b}\right) = P_{\mathrm{RFC}} \left\| \mathbf{c}\left(\mathbf{S}_1\right) \right\|_0 + P_{\mathrm{D}} \sum_{i \in \mathcal{C}} 2^{b_i} \tag{23}$$

$$P_2\left(\mathbf{S}_2\right) = \frac{1}{\rho_{pa}} P_t \left\| \mathbf{r}\left(\mathbf{S}_2\right) \right\|_0 + N_s \frac{\left\| \mathbf{r}\left(\mathbf{S}_2\right) \right\|_0}{N_t} \tag{24}$$

$\boldsymbol{b} = [b_1, b_2, \ldots, b_{NtRF}]^T$表示射频链路上 DAC 的分辨率。

### 2.4 混合预编码优化问题

本文的优化目标为最大化所提出的毫米波混合波束赋形系统的发射端能源效率 $E$，这一指标能很好地表征通信系统在功耗与性能表现之间的权衡，具体而言，能源效率 $E$ 可被定义为频谱效率 $R$ 与发射端系统功耗 $P$ 之比[27]：

$$E = \frac{R}{P} \tag{25}$$

频谱效率由式（12）给出，功耗由式（21）给出，则有：

$$E=\frac{\log_2\left|\mathbf{I}_{N_s}+\frac{\rho}{N_s}\left(\sigma_n^2\mathbf{W}^H\mathbf{W}\right)^{-1}\mathbf{W}^H\mathbf{H}\mathbf{S}_2\mathbf{F}_{\mathrm{ps}}\mathbf{S}_1\boldsymbol{\Delta}\mathbf{F}_{\mathrm{BB}}\mathbf{F}_{\mathrm{BB}}^H\boldsymbol{\Delta}^H\mathbf{S}_1{}^H\mathbf{F}_{\mathrm{ps}}{}^H\mathbf{S}_2{}^H\mathbf{H}^H\mathbf{W}\right|}{P_{\mathrm{con}}+P_1\left(\mathbf{S}_1,\boldsymbol{b}\right)+P_2\left(\mathbf{S}_2\right)} \tag{26}$$

本文主要关注在发射端的预编码架构及算法设计，可以将联合发射端与接收端的优化问题解耦[14]，只考虑发射端的预编码矩阵设计，在假设接收端完美解码的情况下，发射端预编码矩阵的指标可以用信号在信道上的互信息 $I$ 来等效频谱效率：

$$I=\log_2\left|\mathbf{I}_{N_r}+\frac{\rho}{N_s\sigma_n^2}\mathbf{H}\mathbf{S}_2\mathbf{F}_{\mathrm{ps}}\mathbf{S}_1\boldsymbol{\Delta}\mathbf{F}_{\mathrm{BB}}\mathbf{F}_{\mathrm{BB}}^H\boldsymbol{\Delta}^H\mathbf{S}_1{}^H\mathbf{F}_{\mathrm{ps}}{}^H\mathbf{S}_2{}^H\mathbf{H}^H\right| \tag{27}$$

为进一步简化问题，本文考虑在固定 $\mathbf{S}_2$ 的情况下进行求解，也就是将移相器阵列与天线阵列之间的连接关系固定，具体可表述为：

$$\mathbf{S}_2=\begin{bmatrix}\mathbf{I}_{N_{ps}}\\ \mathbf{0}_{(N_t-N_{ps})\times N_{ps}}\end{bmatrix} \tag{28}$$

此时有：

$$P_2\left(\mathbf{S}_2\right)=P_2=\frac{1}{\rho_{pa}}P_tN_{ps}+N_s\frac{N_{ps}}{N_t} \tag{29}$$

综上所述，仅考虑发射端预编码设计的情况下，系统等效能源效率 $\tilde{E}$ 可表示为：

$$\begin{aligned}\tilde{E}&=\frac{I}{P}\\&=\frac{\log_2\left|\mathbf{I}_{N_r}+\frac{\rho}{N_s\sigma_n^2}\mathbf{H}\mathbf{S}_2\mathbf{F}_{\mathrm{ps}}\mathbf{S}_1\boldsymbol{\Delta}\mathbf{F}_{\mathrm{BB}}\mathbf{F}_{\mathrm{BB}}^H\boldsymbol{\Delta}^H\mathbf{S}_1{}^H\mathbf{F}_{\mathrm{ps}}{}^H\mathbf{S}_2{}^H\mathbf{H}^H\right|}{P_{\mathrm{con}}+P_1\left(\mathbf{S}_1,\boldsymbol{b}\right)+P_2}\end{aligned} \tag{30}$$

实际上在图 1 所描述的模拟预编码部分结构中，前置开关阵列还是后置开关阵列的“选择权”均在移相器阵列侧，在所提出的架构中采用的是“单刀多掷”类型的射频开关，这表明每个移相器在前端只能选择一个射频链路进行连接，在其后端只能选择一个天线阵列元进行连接，对于前置开关矩阵 $\mathbf{S}_1$ 来说有：

$$\left\|\mathbf{S}_{1i,:}\right\|_0=1,\quad i=1,2\ldots,N_{ps} \tag{31}$$

其中 $\mathbf{S}_{1i,:}$表示 S1 的第 $i$ 行。

最大化等效能源效率优化问题可表述为：

$$\max_{\mathbf{F}_{\mathrm{BB}},\mathbf{S}_1,\mathbf{F}_{\mathrm{ps}},\boldsymbol{b}} \frac{\log_2\left|\mathbf{I}_{N_r}+\frac{\rho}{N_s\sigma_n^2}\mathbf{H}\mathbf{S}_2\mathbf{F}_{\mathrm{ps}}\mathbf{S}_1\boldsymbol{\Delta}\mathbf{F}_{\mathrm{BB}}\mathbf{F}_{\mathrm{BB}}^H\boldsymbol{\Delta}^H\mathbf{S}_1{}^H\mathbf{F}_{\mathrm{ps}}{}^H\mathbf{S}_2{}^H\mathbf{H}^H\right|}{P_{\mathrm{con}}+P_1\left(\mathbf{S}_1,\boldsymbol{b}\right)+P_2}$$

$$\text{s.t.}\begin{cases}\mathbf{S}_1\in\mathcal{B}^{N_{ps}\times N_{tRF}}\\ \left\|\mathbf{S}_{1i,:}\right\|_0=1, i=1,2\ldots,N_{ps}\\ b_i\in\left[b_{min},b_{max}\right], i=1,2,\ldots,N_{tRF}\\ \mathbf{F}_{\mathrm{ps}}\in\mathcal{F}\\ \left\|\mathbf{F}_{\mathrm{RF}}\mathbf{F}_{\mathrm{BB}}\right\|_F^2=N_s\frac{N_{ps}}{N_t}\end{cases} \tag{32}$$

其中$\mathcal{B}$表示二元“01”矩阵的集合空间，$\mathcal{F}$表示矩阵 $\mathbf{F}_{\mathrm{ps}}$ 的可能取值集合空间，其对角线元素的相位满足式（7）中的约束，并满足恒模约束。$b_i$ 为 $\boldsymbol{b}$ 中的第 $i$ 个元素，由式（5）可知，矩阵 $\boldsymbol{\Delta}$ 可表示为 $\boldsymbol{b}$ 的函数。

## 3 基于块坐标下降的混合预编码算法

值得注意的是，优化问题（32）具有分式形式的目标函数，分母为离散变量的函数，同时优化变量具有非凸二元约束，这使得该问题的求解变得具有挑战性。由此，本文针对该问题提出了一种基于块坐标下降（Block Coordinate Descent, BCD）思想的混合预编码算法。

BCD 算法的核心思想是将优化变量分为多个块，每次迭代中只优化一个块的变量，保持其他块的变量固定。将复杂优化问题转换为一系列子问题，问题（32）中的各变量相互独立，较为适合使用 BCD 的思想，进行迭代求解。

该优化问题的自变量主要有 $\mathbf{F}_{\mathrm{BB}}$、$\mathbf{S}_1$、$\mathbf{F}_{\mathrm{ps}}$ 和 $\boldsymbol{b}$。需要考虑如何合理划分块变量从而使问题简化，重点是如何将 $\mathbf{S}_1$ 从分子中分离。文献[28]考虑了动态子阵列结构下的混合预编码设计问题，以最大化系统频谱效率为目标函数，仿真结果表明，该架构下以频谱效率为优化目标函数，求得的模拟预编码矩阵中的非零行数总是为 $N_s$，表明只使用了 $N_s$ 条射频链路，为保障多流通信，有 $N_s \leqslant N_{tRF}$，故 $N_s$ 为其下界。结合所提出的架构，若以最大化频谱效率为目标函数求解 $\mathbf{S}_1$，则$\|\mathbf{c}(\mathbf{S}_1)\|_0 = N_s$。故可将变量分为两组：$\mathbf{F}_{\mathrm{BB}}$、$\mathbf{S}_1$ 与 $\mathbf{F}_{\mathrm{ps}}$ 为一组，以最大化发射端互信息为目标函数，$\boldsymbol{b}$ 单独为一组，求解上一组变量固定下的问题，此时问题（32）可分为两个子问题：

$$\max_{\mathbf{F}_{\mathrm{BB}},\mathbf{S}_1,\mathbf{F}_{\mathrm{ps}}} \log_2\left|\mathbf{I}_{N_r}+\frac{\rho}{N_s\sigma_n^2}\mathbf{H}\mathbf{S}_2\mathbf{F}_{\mathrm{ps}}\mathbf{S}_1\boldsymbol{\Delta}\mathbf{F}_{\mathrm{BB}}\mathbf{F}_{\mathrm{BB}}^H\boldsymbol{\Delta}^H\mathbf{S}_1{}^H\mathbf{F}_{\mathrm{ps}}{}^H\mathbf{S}_2{}^H\mathbf{H}^H\right|$$

$$\text{s.t.}\begin{cases}\mathbf{S}_1\in\mathcal{B}^{N_{ps}\times N_{tRF}}\\ \left\|\mathbf{S}_{1i,:}\right\|_0=1, i=1,2\ldots,N_{ps}\\ \mathbf{F}_{\mathrm{ps}}\in\mathcal{F}\\ \left\|\mathbf{F}_{\mathrm{RF}}\mathbf{F}_{\mathrm{BB}}\right\|_F^2=N_s\frac{N_{ps}}{N_t}\end{cases} \tag{33}$$

以及

$$\max_{\boldsymbol{b}} \frac{\log_2\left|\mathbf{I}_{N_r}+\frac{\rho}{N_s\sigma_n^2}\mathbf{H}\mathbf{S}_2\mathbf{F}_{\text{ps}}\mathbf{S}_1\boldsymbol{\Delta}\mathbf{F}_{\text{BB}}\mathbf{F}_{\text{BB}}^H\boldsymbol{\Delta}^H\mathbf{S}_1{}^H\mathbf{F}_{\text{ps}}{}^H\mathbf{S}_2{}^H\mathbf{H}^H\right|}{P_{\text{con}}+P_1\left(\mathbf{S}_1,\boldsymbol{b}\right)+P_2} \tag{34}$$
$$\text{s.t.}\, b_i\in\left[b_{min},b_{max}\right], i=1,2,\ldots,N_{tRF}$$

对于子问题（33），其可以等价为最小化混合预编码矩阵与最佳预编码矩阵 $\mathbf{F}_{\text{opt}}$ 之间的欧式距离的优化问题，即：

$$\min_{\mathbf{F}_{\text{BB}},\mathbf{S}_1,\mathbf{F}_{\text{ps}}} \left\|\mathbf{F}_{\text{opt}}-\mathbf{S}_2\mathbf{F}_{\text{ps}}\mathbf{S}_1\boldsymbol{\Delta}\mathbf{F}_{\text{BB}}\right\|_F^2$$
$$\text{s.t.}\quad \begin{array}{c} \mathbf{S}_1\in\mathcal{B}^{N_{ps}\times N_{tRF}} \\ \left\|\mathbf{S}_{1i,:}\right\|_0=1, i=1,2\ldots,N_{ps} \\ \mathbf{F}_{\text{ps}}\in\mathcal{F} \\ \left\|\mathbf{F}_{\text{RF}}\mathbf{F}_{\text{BB}}\right\|_F^2=N_s\frac{N_{ps}}{N_t} \end{array} \tag{35}$$

这里的 $\mathbf{F}_{\text{opt}}$ 可通过对 CSI 矩阵 $\mathbf{H}$ 进行 SVD 分解得到。如此，问题（32）可等价为子问题（34）和子问题（35）的迭代求解，基于 BCD 的迭代求解算法具体如算法 1 所示：

---

**算法 1** 基于 BCD 的问题（32）的求解算法

---

输入：$\mathbf{H}, q, N_t, N_{ps}, N_{\text{s}}, \mathbf{S}_2$, SNR

输出：$\mathbf{F}_{\text{BB}}, \mathbf{S}_1, \mathbf{F}_{\text{ps}}, \boldsymbol{b}$

**1** 初始化，设迭代次数 $k=0$，随机生成变量 $\boldsymbol{b}$

**2 while** 未收敛 **or** $k < N_{iter}$ **do**

**3** 　固定变量 $\boldsymbol{b}$，求解子问题（35），得到块变量 $\mathbf{F}_{\text{BB}}$、$\mathbf{S}_1$ 与 $\mathbf{F}_{\text{ps}}$ 的解

**4** 　固定变量 $\mathbf{F}_{\text{BB}}$、$\mathbf{S}_1$ 与 $\mathbf{F}_{\text{ps}}$，求解子问题（34），得到块变量 $\boldsymbol{b}$ 的解

**5 return** $\mathbf{F}_{\text{BB}}, \mathbf{S}_1, \mathbf{F}_{\text{ps}}, \boldsymbol{b}$

---

考虑子问题（34）的求解，自变量 $\boldsymbol{b}$ 的取值是离散的，这意味着应用搜索的方法来求解是可行的，特别地，若应用遍历的方法求解，由于 $\boldsymbol{b}$ 共有 $N_{tRF}$ 个元素，一共只需要进行$(b_{max} - b_{min} + 1)\times N_{tRF}$ 次数的计算，从中比较各个情况下的能源效率，最后以最高的能源效率下的取值作为求解结果。这对于问题的求解来说是可取的。故使用遍历的方法来求解子问题（34）。

对于子问题（35）的求解，由于变量的相关非凸约束与矩阵的耦合，直接求解该问题仍是比较困难的，因此本节考虑使用交替最小化的思想进行求解，即交替固定某一变量来优化另一个。在三个变量中，$\mathbf{F}_{\text{BB}}$ 约束最少，可优先求解，在分析时可将 $\boldsymbol{\Delta}$ 近似为单位矩阵。固定 $\mathbf{S}_1$ 与 $\mathbf{F}_{\text{ps}}$，同时对其施加半酉约束以便求解，问题（35）可表述为：

$$\min_{\mathbf{F}_{\text{BB}}} \left\|\mathbf{F}_{\text{opt}}-\mathbf{S}_2\mathbf{F}_{\text{ps}}\mathbf{S}_1\mathbf{F}_{\text{BB}}\right\|_F^2$$
$$\text{s.t.}\quad \mathbf{F}_{\text{BB}}^H\mathbf{F}_{\text{BB}}=\mathbf{I}_{N_s} \tag{36}$$

该问题是一个正交普鲁克问题，解为[15]:

$$\begin{gathered}\mathbf{F}_{\mathrm{BB}} = \mathbf{V}(:,1:N_s)\mathbf{U}^H\mathbf{F}_{\mathrm{opt}}^H \\ \mathbf{S}_2\mathbf{F}_{\mathrm{ps}}\mathbf{S}_1 = \mathbf{U}\boldsymbol{\Sigma}\mathbf{V}^H\end{gathered} \tag{37}$$

考虑在固定 $\mathbf{F}_{\mathrm{BB}}$ 的情况下求解 $\mathbf{S}_1$ 与 $\mathbf{F}_{\mathrm{ps}}$，此时优化问题为：

$$\begin{gathered}\min_{\mathbf{S}_1,\mathbf{F}_{\mathrm{ps}}} \left\|\mathbf{F}_{\mathrm{opt}} - \mathbf{S}_2\mathbf{F}_{\mathrm{ps}}\mathbf{S}_1\mathbf{F}_{\mathrm{BB}}\right\|_F^2 \\ \mathbf{S}_1 \in \mathcal{B}^{N_{ps}\times N_{tRF}} \\ \text{s.t.} \left\|\mathbf{S}_{1i,:}\right\|_0 = 1, i = 1,2\ldots,N_{ps} \\ \mathbf{F}_{\mathrm{ps}} \in \mathcal{F}\end{gathered} \tag{38}$$

其中 $\mathbf{F}_{\mathrm{RF}} = \mathbf{S}_2\mathbf{F}_{\mathrm{ps}}\mathbf{S}_1$，注意到：

$$\left\|\mathbf{F}_{\mathrm{opt}} - \mathbf{S}_2\mathbf{F}_{\mathrm{ps}}\mathbf{S}_1\mathbf{F}_{\mathrm{BB}}\right\|_F^2 \approx \mathrm{Tr}\left(\mathbf{F}_{\mathrm{opt}}\mathbf{F}_{\mathrm{opt}}^H\right) + \mathrm{Tr}\left(\mathbf{F}_{\mathrm{RF}}\mathbf{F}_{\mathrm{RF}}^H\right) - 2\,\mathrm{Tr}\left(\mathrm{Re}\left(\mathbf{F}_{\mathrm{RF}}\mathbf{F}_{\mathrm{BB}}\mathbf{F}_{\mathrm{opt}}^H\right)\right) \tag{39}$$

多项式（39）的第一项与第二项均为定值，故最小化式（39）等价于最大化其最后一项，即等价于如下问题：

$$\begin{gathered}\max_{\mathbf{S}_1,\mathbf{F}_{\mathrm{ps}}} \mathrm{Tr}\left(\mathrm{Re}\left(\mathbf{S}_2\mathbf{F}_{\mathrm{ps}}\mathbf{S}_1\mathbf{F}_{\mathrm{BB}}\mathbf{F}_{\mathrm{opt}}^H\right)\right) \\ \mathbf{S}_1 \in \mathcal{B}^{N_{ps}\times N_{tRF}} \\ \text{s.t.} \left\|\mathbf{S}_{1i,:}\right\|_0 = 1, i = 1,2\ldots,N_{ps} \\ \mathbf{F}_{\mathrm{ps}} \in \mathcal{F}\end{gathered} \tag{40}$$

令 $\mathbf{F} = \mathbf{F}_{\mathrm{BB}}\mathbf{F}_{\mathrm{opt}}^{\mathrm{H}} = [\mathbf{F}_1,\mathbf{F}_2,\ldots,\mathbf{F}_{Nt}]$，$\mathbf{S}_{1i,:}$表示 $\mathbf{S}_1$ 的第 $i$ 行，$\mathbf{S}_{2i,j}$ 为 $\mathbf{S}_2$ 第 $i$ 行第 $j$ 列的元素，引入 $\mathbf{S}_{xi} = e^{j\phi i}\mathbf{S}_{1i,:}$，问题（40）的目标函数可化为：

$$\sum_{i=1}^{N_{ps}}\sum_{j=1}^{N_t}\mathrm{Re}\left(\mathrm{S}_{2j,i}\mathbf{S}_{xi}\mathbf{F}_j\right) \tag{41}$$

易知对于任意的 $i$，$\mathbf{S}_{xi}$ 是独立的，故可将其分解为 $N_{ps}$ 个相同的子问题：

$$\begin{gathered}\max_{\mathbf{S}_{xi}} \mathbf{S}_{xi}\sum_{j=1}^{N_t}\mathrm{Re}\left(\mathrm{S}_{2j,i}\mathbf{F}_j\right) \\ \left\|\mathbf{S}_{1i,:}\right\|_0 = 1 \\ s.t. \quad \Phi_i \in \left\{0,\frac{2\pi}{2^q},\ldots,\frac{2\pi\left(2^q-1\right)}{2^q}\right\}\end{gathered} \tag{42}$$

令 $\mathcal{M}_i = \{m|\mathrm{S}_{2m,i} = 1, m = 1, 2,\ldots, N_t\}$，则有：

$$\mathbf{S}_{xi}\sum_{j=1}^{N_t}\mathrm{Re}\left(\mathrm{S}_{2j,i}\mathbf{F}_j\right) = \mathbf{S}_{xi}\,\mathrm{Re}\left(\sum_{j\in\mathcal{M}_i}\mathbf{F}_j\right) \tag{43}$$

显然 $\mathbf{S}_{xi}$ 是一个元素为 $e^{j\Phi i}$，其余元素都为零的稀疏行向量，可用贪心的思想求解问题（42）：可先找到 $\Sigma\mathbf{F}_j$ 中绝对值最大的元素，记索引为 $n$，要使 $\mathbf{S}_{xi}\mathrm{Re}(\Sigma\mathbf{F}_j)$最大，则 $\Phi_i$ 应满足：

$$\Phi_i = 2\pi - \mathrm{Arg}\left(\left[\sum_{j\in\mathcal{M}}\mathbf{F}_j\right]_n\right) \tag{44}$$

由于 $\Phi_i$ 有如问题（42）中的量化相位约束，故需要在相位集合中选择与式（44）中最为接近的一项作为 $\Phi_i$ 的取值，记为 $\Phi_i^{'}$，此时 $\mathbf{S}_{xi}$ 可表示为：

$$\left[\mathbf{S}_{xi}\right]_n = e^{j\widehat{\Phi}_i} \tag{45}$$

所有形如问题（42）被求解后，可以得到 $\mathbf{S}_x$，从中可还原出 $\mathbf{S}_1$ 与 $\mathbf{F}_{\mathrm{ps}}$。完成问题（38）的求解后，可将求出的变量值固定，继续问题（36）的求解，直至满足跳出迭代的条件。迭代结束后，需对 $\mathbf{F}_{\mathrm{BB}}$ 进行归一化以满足功率约束：

$$\mathbf{F}_{\mathrm{BB}} \leftarrow \frac{\sqrt{N_s N_{ps} / N_t}}{\left\|\mathbf{S}_2\mathbf{F}_{\mathrm{ps}}\mathbf{S}_1\mathbf{F}_{\mathrm{BB}}\right\|_F}\mathbf{F}_{\mathrm{BB}} \tag{46}$$

综上所述，目标优化问题的求解流程如图 2 所示。

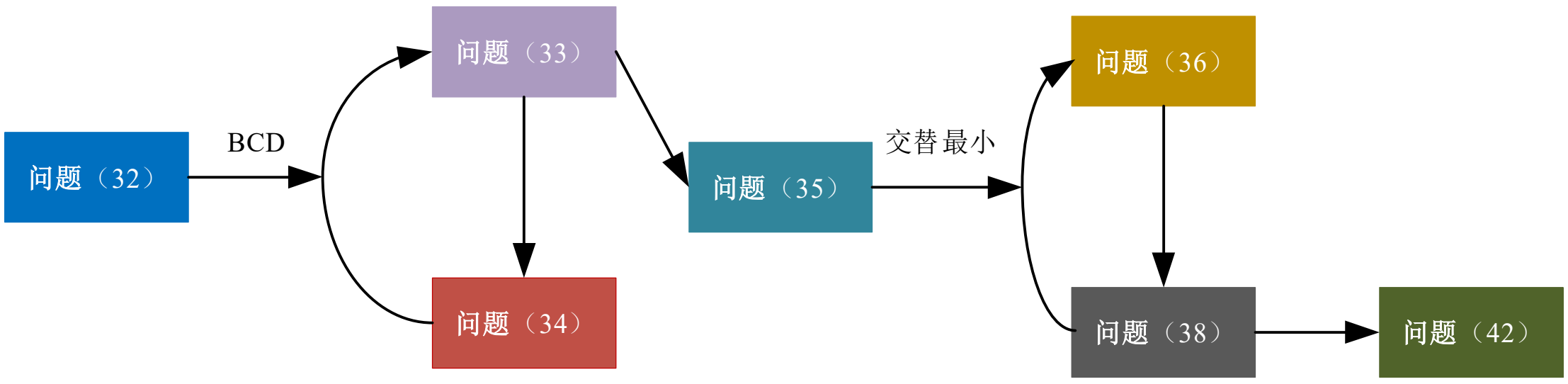

图 2　　基于块坐标下降的混合预编码算法流程

在所提出的基于块坐标下降的混合预编码算法中，通过迭代优化数字预编码矩阵、模拟预编码矩阵以及 DAC 的分辨率，降低了全局优化问题的复杂度。具体而言，$\mathbf{F}_{\mathrm{BB}}$ 的优化基于 SVD 计算，其计算复杂度为 $O(N_t N_s^2)$，模拟预编码矩阵的优化包括 $\mathbf{S}_1$ 与 $\mathbf{F}_{\mathrm{ps}}$ 的求解，利用稀疏矩阵的贪婪搜索的思想，总体的计算复杂度为 $O(N_{ps} N_t)$，DAC 的分辨率 $\boldsymbol{b}$ 通过遍历搜索求解，复杂度为 $O((b_{max} - b_{min} + 1) \times N_{tRF})$，则交替最小化求解问题（35）的复杂度为 $O(N_{iter2} \times (N_t N_s^2 + N_{ps} N_t))$。综上，单次迭代的总体复杂度可表示为 $O(N_{iter2} \times (N_t N_s^2 + N_{ps} N_t) + (b_{max} - b_{min} + 1) \times N_{tRF})$。设总迭代次数为 $N_{iter1}$，则所提出算法整体的计算复杂度可表示为 $O(N_{iter1} \times (N_{iter2} \times (N_t N_s^2 + N_{ps} N_t) + (b_{max} - b_{min} + 1) \times N_{tRF}))$。

# 4　仿真结果与分析

本节采用开源毫米波信道仿真软件 NYUSIM[29-30]来生成信道矩阵 $\mathbf{H}$，并分别在完整 CSI 与部分 CSI 下进行系统频谱效率以及能源效率的数值仿真，并与一些具有代表性的基线算法作对比。

## 4.1　仿真参数设置

在所提出的混合波束赋形系统中，工作频率被设置为 28GHz，发射端天线数 $N_t$ 为 64，接收端天线数 $N_r$ 为 16。发射端的射频链路数 $N_{tRF}$ 为 4，数据流个数 $N_s$ 为 1 或 2。假定接收端使用全分辨率的全数字波束赋形，发射端移相器个数 $N_{ps}$ 为 50，量化比特数 $q = 4$，BCD 算法的最大迭代次数 $N_{iter1}$ 设置为 10，交替最小算法的最大迭代次数 $N_{iter2}$ 设置为 20，信噪比（Signal-to-Noise Ratio, SNR）设置为-20dB 至 20dB。

硬件功耗参数方面参考了文献[31]的结论，如表 1 所示：

表 1　　混合波束赋形硬件功耗参数设置

| 说明 | 设置值 |
| --- | --- |
| DAC 每比特功耗 | 0.39 mW |
| 射频转换器功耗 | 40 mW |
| 射频开关功耗 | 5 mW |
| 可调移相器功耗 | 30 mW |
| 天线阵列元功耗 | 20 mW |
| 基带编码器功耗 | 200 mW |
| 功率放大器转换效率 | 30% |
| DAC 最小分辨率 | 4 bit |
| DAC 最大分辨率 | 16 bit |

NYUSIM 软件的相关参数设置见表 2：

表 2　　NYUSIM 主要仿真参数设置

| 参数 | 设置值 |
| --- | --- |
| 场景 | UMi |
| 载波频率 | 28GHz |
| 环境 | LOS |
| 极化 | Co-Pol |
| 接收端位置数量 | 50 |
| 发射端/接收端天线阵列类型 | URA |
| 发送端天线阵列元数量（Nt） | 64 |
| 接收端天线阵列元数量（Nr） | 16 |
| 发射端/接收端天线方位半功率波束宽 | 180° |

同时，本文选取了一些具有代表性的基线算法作为对比：全数字波束赋形（FD）、全连接架构（FC）下基于 OMP 的预编码算法[14]以及动态子阵列架构（DSA）下基于 AltMin 的预编码算法[16]。

**4.2　完整 CSI 下的仿真结果**

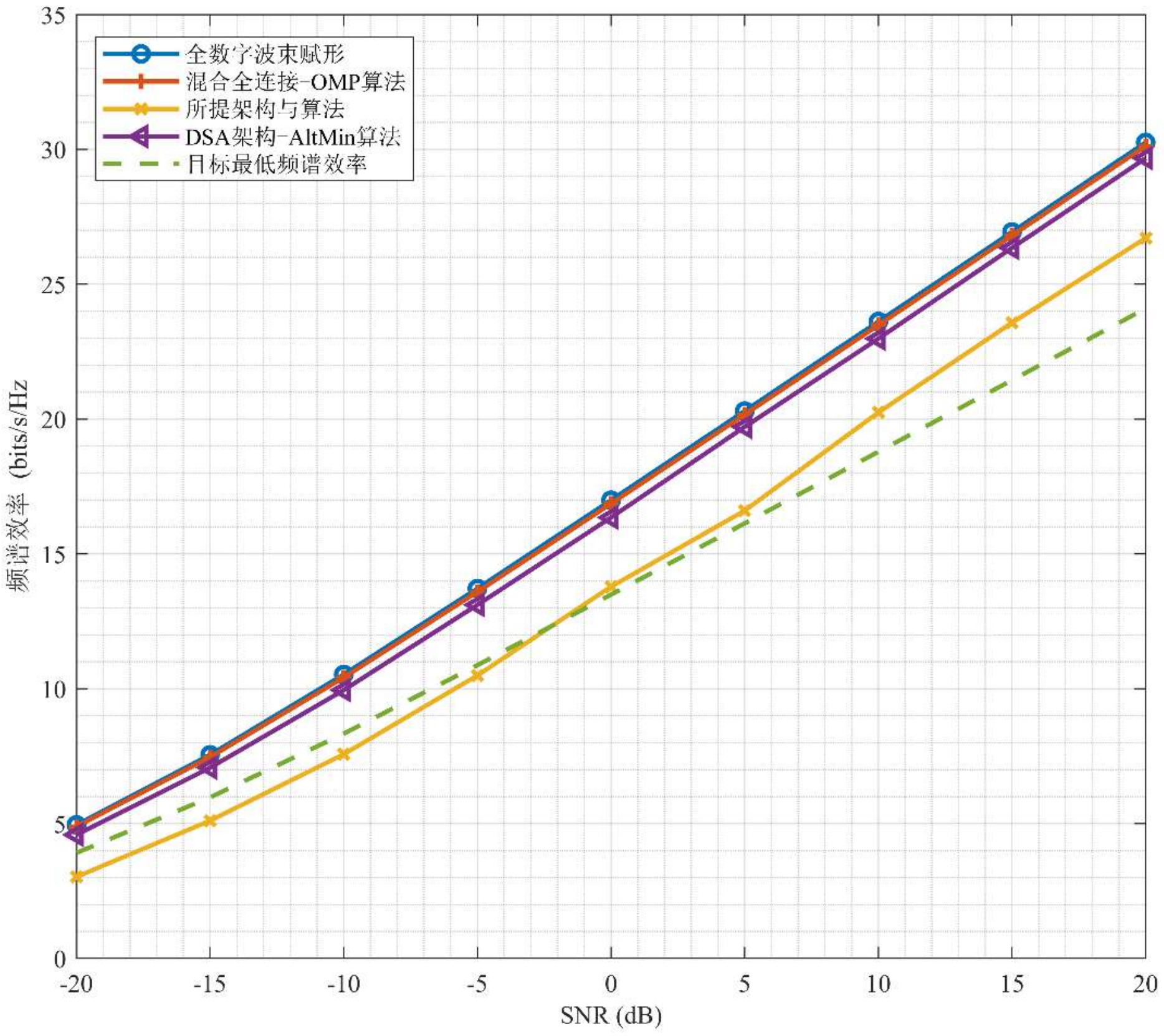

图 3　　不同架构与预编码算法的频谱效率与 SNR 之间的关系

图 3 表明，当 SNR 大于 0dB 时，所提出的方法的频谱效率基本高于最低目标频谱效率，该指标被设定为全连接架构下的基于 OMP 的预编码算法所达成频谱效率的 80%，在 SNR 为 2dB 到 6dB 左右时，其频谱效率与全连接架构与全数字波束赋形架构的差值在 3.5 bits/s/Hz 左右，相较于全连接架构，频谱效率在损失 17%左右，仍满足最低目标 20%。

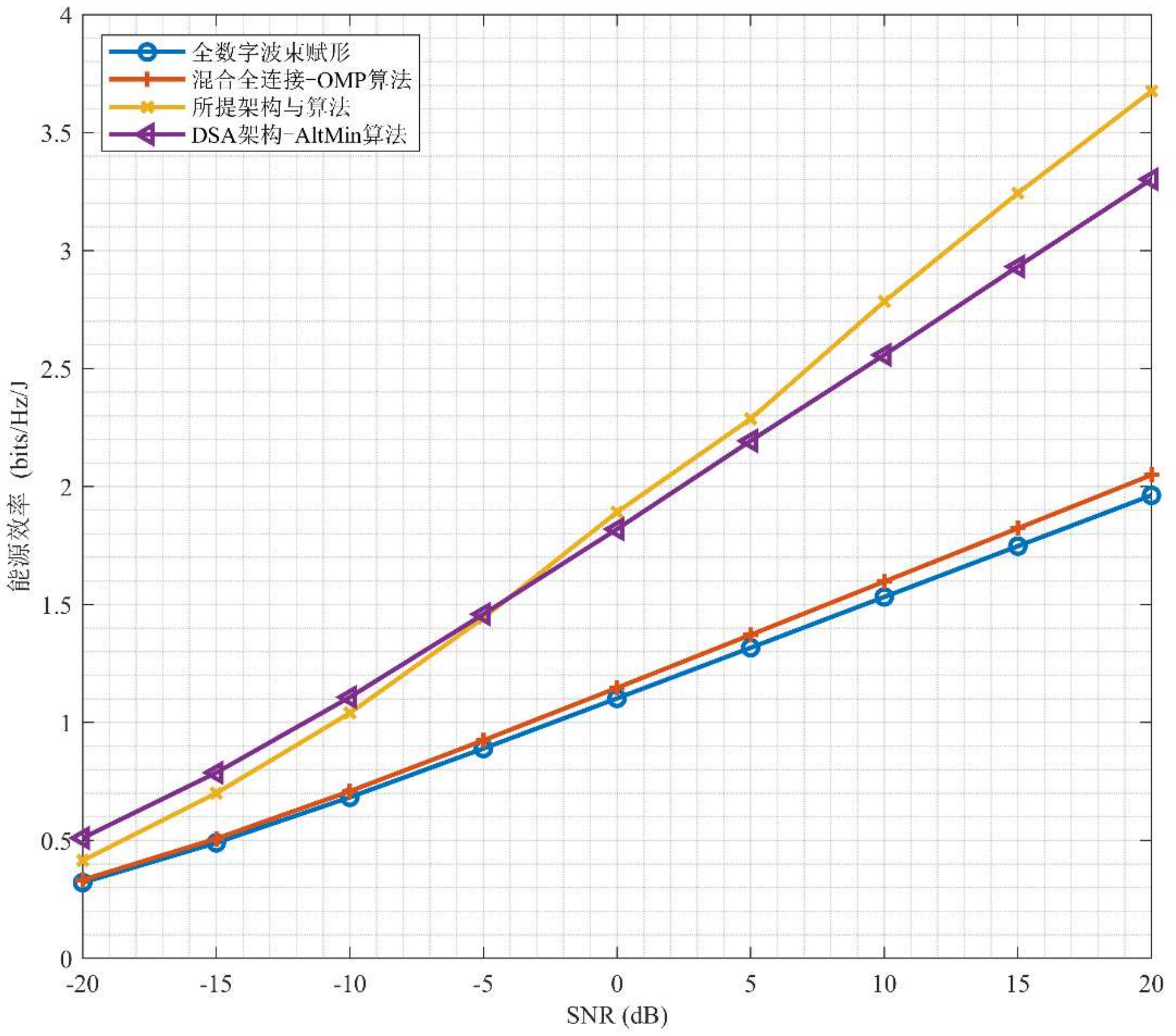

图 4　　不同架构与预编码算法的能源效率与 SNR 之间的关系

图 4 展示了各架构与算法在不同 SNR 下的能源效率，在所有 SNR 下，本文所提出的架构与算法在能源效率上表现最好，其次是使用 AltMin 预编码算法的动态子阵列架构，而这比较接近。这是因为这两种架构都能根据 $N_s$ 的大小合理利用射频链路，同时本文提出的架构与算法又近一步优化了 DAC 的分辨率，并通过两处开关减少了使用移相器的数量，这些都提升了系统的能源效率。

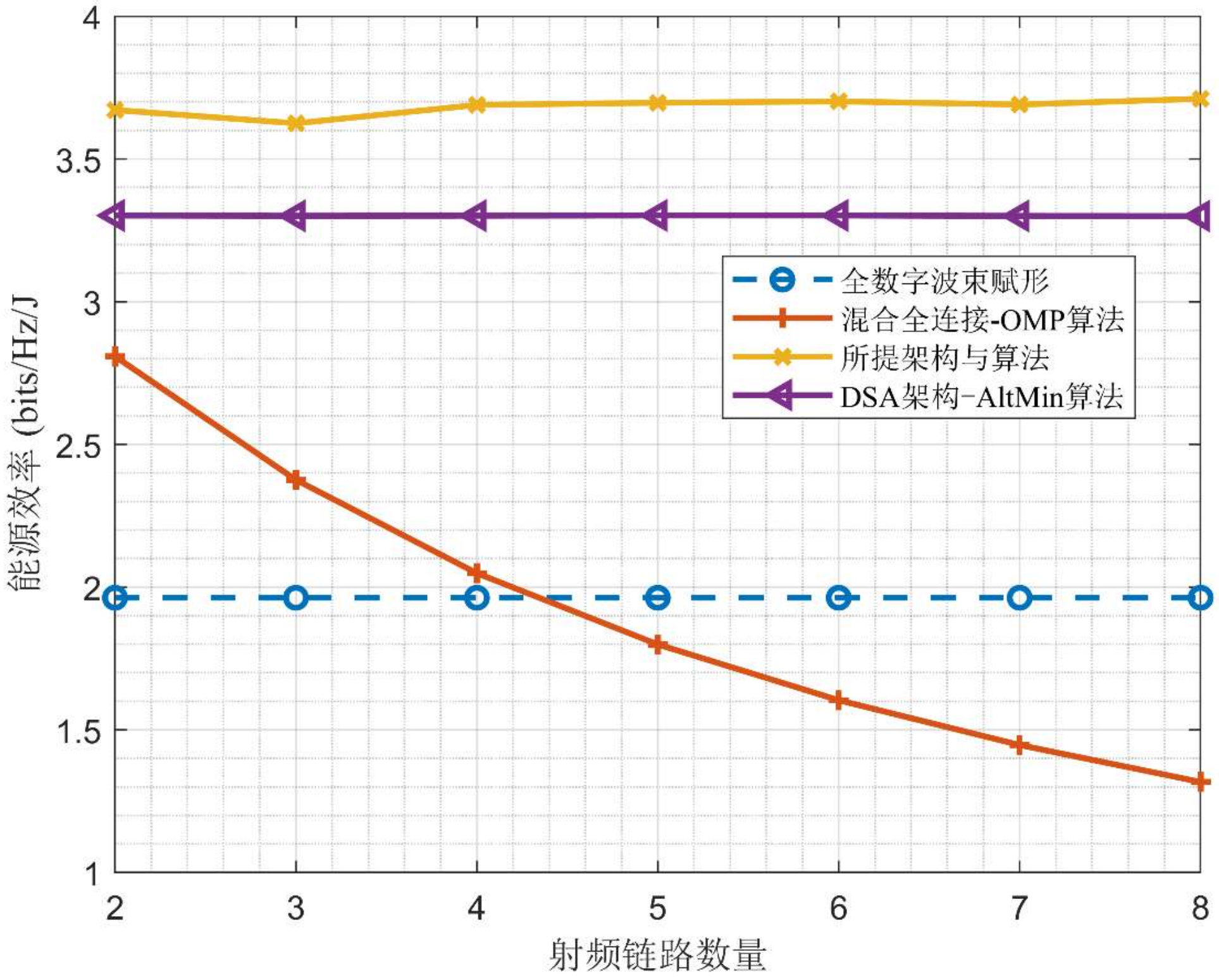

图 5　　不同架构与预编码算法的能源效率与射频链路数量 $N_{tRF}$ 之间的关系

从图 5 中展示的能源效率与 $N_{tRF}$ 之间的关系可以看出，全连接架构由于 $N_{tRF}$ 的增加，使用的移相器数也随之增加，大幅提升了系统功耗，但频谱效率的提升却有限，这导致了能源效率的不断降低。而本文提出的架构在各种情况下的能源效率总是最高，这是因为该架构能根据数据流大小 $N_s$ 来动态连接射频链路，多余的射频链路不会被激活，并且该架构选择了一定数量的天线子阵列进行连接，未被连接的天线的功耗可被忽略。这些均体现了所提出架构的节能性。

值得注意的是，实际信道测量中得到的 $\mathbf{H}$ 并不一定能完美反映信道状态，往往是带有偏差，需要评估不准确的 CSI 矩阵对所提出的架构与算法的影响。具体而言，不准确的 CSI 矩阵可以表示为[8]：

$$\widehat{\mathbf{H}} = \xi \mathbf{H} + \sqrt{1-\xi^2}\mathbf{E} \tag{47}$$

$\mathbf{H}$ 表示实际的 CSI 矩阵，$\xi \in [0,1]$表示 CSI 的准确率，$\mathbf{E}$ 为误差矩阵，用来表示估计过程中的误差，其每个元素都互相独立且遵循均值为 0，方差为 1 的复高斯分布。

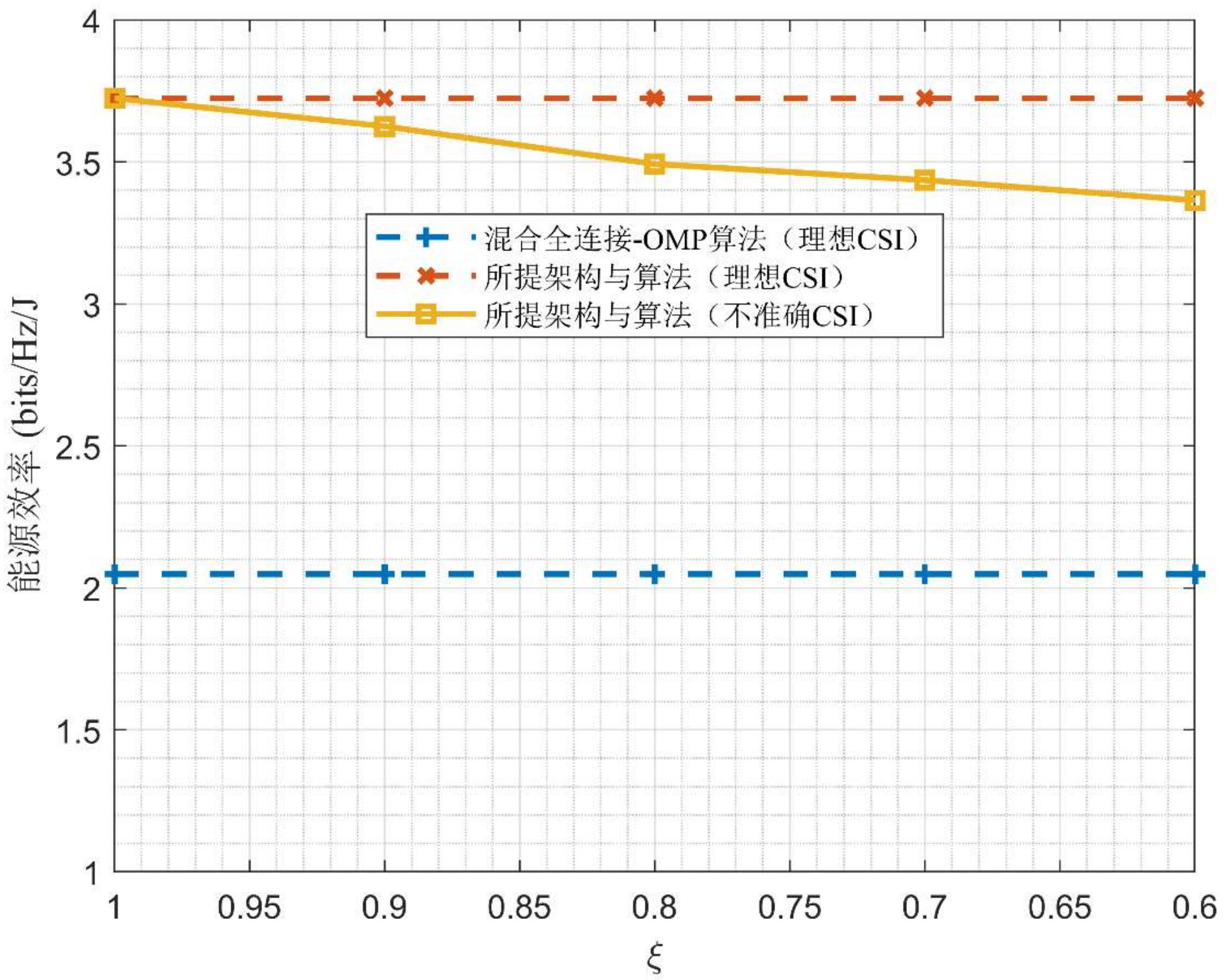

图 6　　不同架构与预编码算法的能源效率与估计准确率 $\xi$ 之间的关系

图 6 显示了 SNR = 20dB 下，各方案在能源效率上的表现。随着 $\xi$ 的增加，所提出的预编码算法的能源效率也逐渐降低，这主要是由频谱效率的降低而导致的。$\xi = 0.6$ 时，所提出的算法的能源效率在 3.4bits/Hz/J 左右，而完美 CSI 时的能源效率在 3.7bits/Hz/J 左右，能源效率损失在 8%左右，表明所提出的预编码算法对 CSI 的估计误差具有较强的鲁棒性。

### 4.3　部分 CSI 下的仿真结果

实际中，完整的 CSI 较难获得，发射端能获得的通常是部分的 CSI，往往为毫米波信道的多径路径增益幅度与相较于发射端天线阵列的方向。目前的仿真结果或是预编码算法设计主要是在完整 CSI 下进行的，有必要验证所提出的算法在部分 CSI 下的有效性。

在本文考虑的场景下，部分 CSI 即为路径增益的幅度与多径的角度，在信道模型中即为式（13）中的 $\mathbf{\Lambda}$ 与 $\mathbf{A}_t$。对于预编码矩阵的求解，只需将 $\mathbf{H}$ 替换为 $\mathbf{\Lambda}\mathbf{A}_t^H$，并稍微调整相关固定参数即可。

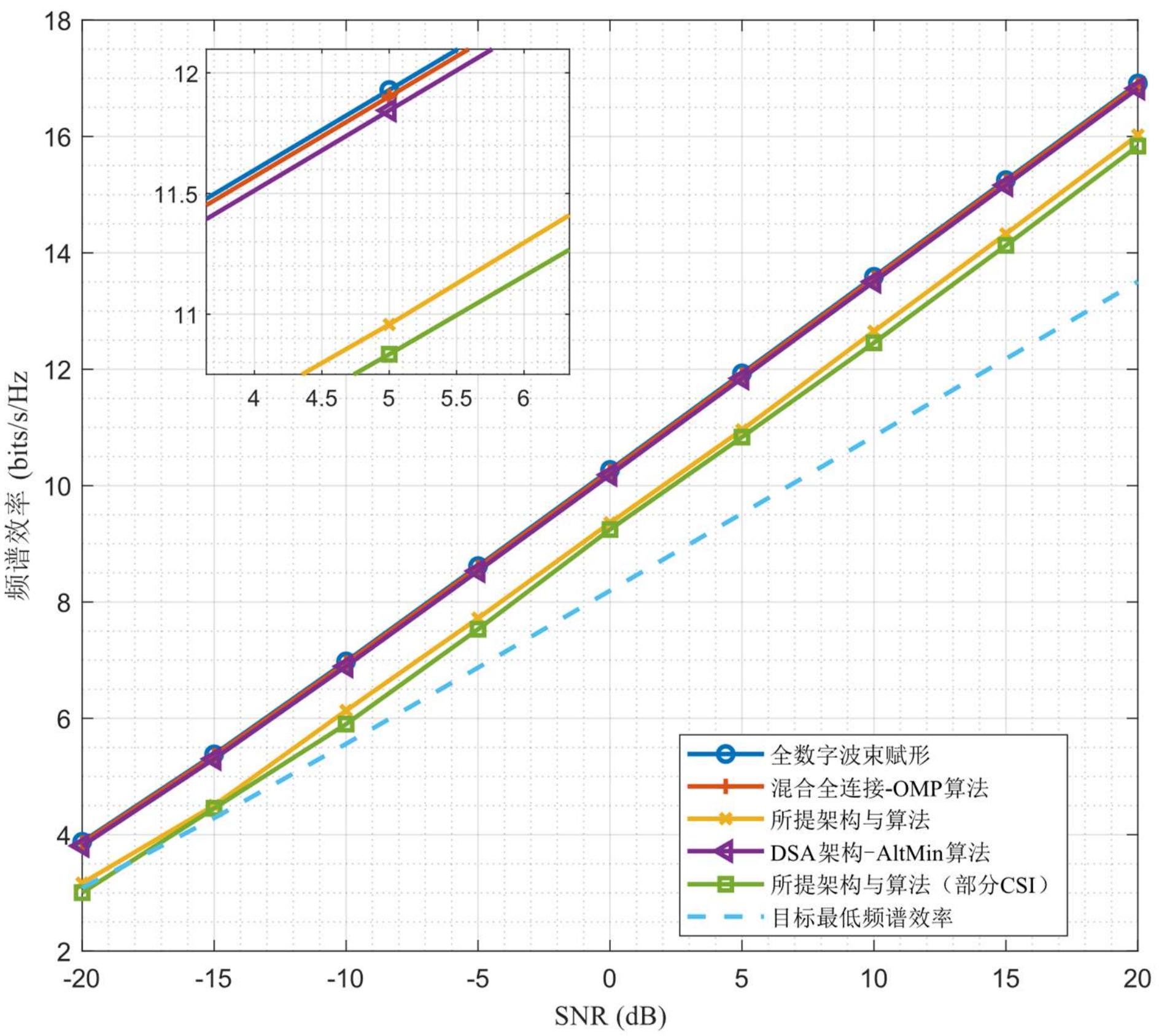

图 7　　部分 CSI 下 $N_s = 1$ 时不同架构与预编码算法的频谱效率与 SNR 之间的关系

如图 7 所示，所提出的预编码算法在部分 CSI 的频谱效率均高于图中的最低目标频谱效率，与在完整 CSI 下所提出的算法达到的频谱效率相比，差值在 0.2bits/s/Hz 左右，损失较小。

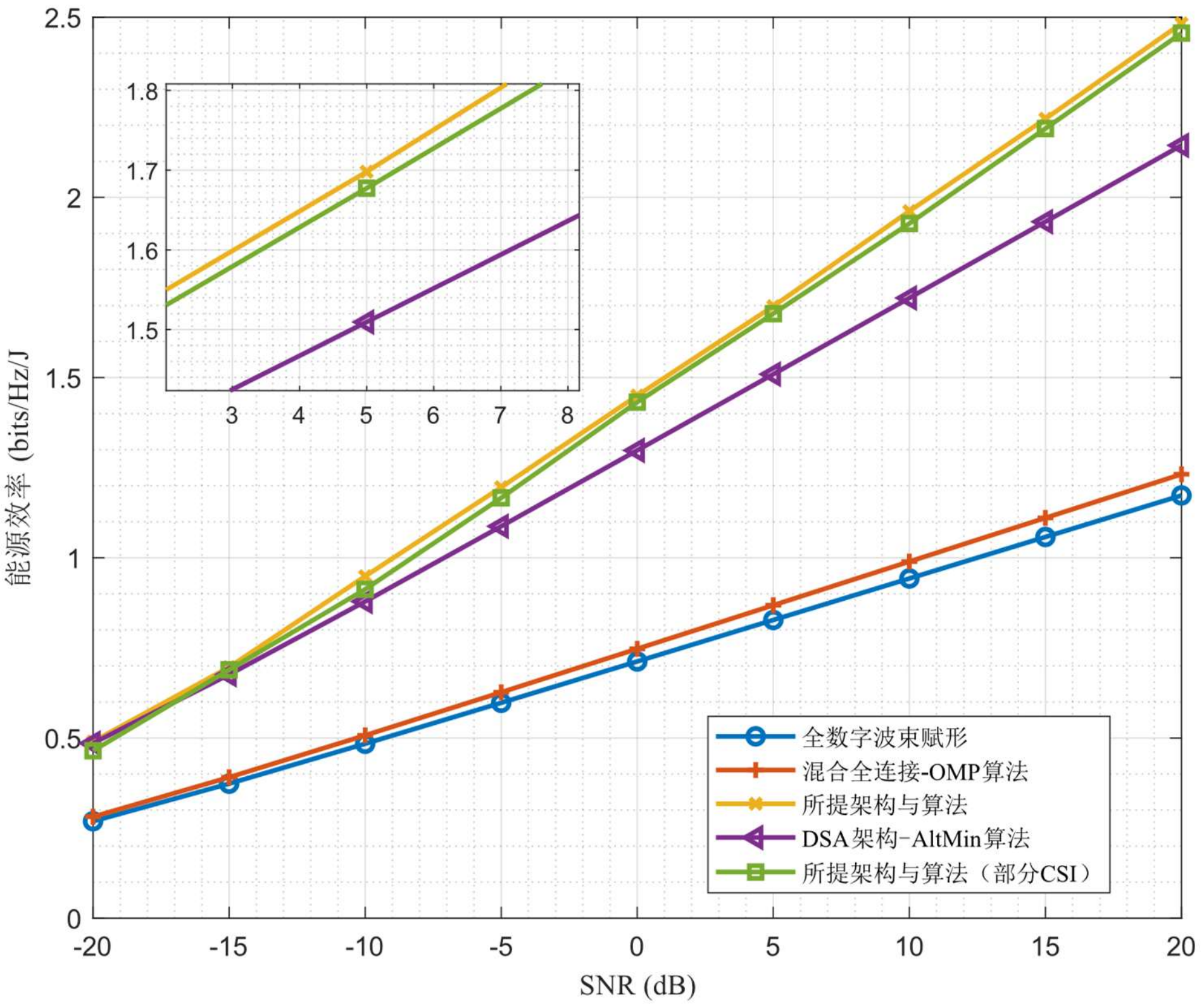

图 8　部分 CSI 下 $N_s = 1$ 时不同架构与预编码算法的能源效率与 SNR 之间的关系

从图 8 中的数据可以看出，各 SNR 下所提出的算法在部分 CSI 的情况下得到的能源效率略低于完整 CSI 下的结果，但差值只有 0.1bits/Hz/J，且结果仍明显高于其余基线算法（完整 CSI 下）的表现。

上述仿真结果表明本文所提出的混合波束赋形架构在大部分情况都具有最高的能源效率，且频谱效率相较于全连接架构的损失较小，有效证明了所提出架构在节省系统功耗上的优越性和实际应用中的有效性。

## 5　结束语

本文针对混合波束赋形的低能源效率与高硬件复杂度问题，提出了一种新型的低功耗混合波束赋形架构，在移相器网络的前后两端引入开关网络，实现了射频链路与移相器以及天线之间的动态连接，同时通过构建包含频谱效率和系统功耗的能源效率模型，提出了基于 BCD 思想的混合预编码算法，最后利用由 NYUSIM 软件生成的毫米波信道矩阵进行了仿真验证。仿真结果证明了多种毫米波通信场景下所提架构及算法在能源效率上的优越性。

参考文献：

[1] RAPPAPORT T S, SUN S, MAYZUS R, et al. Millimeter Wave Mobile Communications for 5G Cellular: It Will Work![J]. IEEE Access, 2013, 1: 335-349.
[2] LU N, CHENG N, ZHANG N, et al. Connected Vehicles: Solutions and Challenges[J]. IEEE Internet of Things Journal, 2014, 1(4): 289-299.
[3] CHEN W, LIN X, LEE J, et al. 5G-Advanced Toward 6G: Past, Present, and Future[J]. IEEE Journal on Selected Areas in Communications, 2023, 41(6): 1592-1619.
[4] IMT-2030(6G)推进组. 6G 典型场景和关键能力白皮书[R]. 2022.
[5] LARSSON E G, EDFORS O, TUFVESSON F, et al. Massive MIMO for next generation wireless systems[J]. IEEE Communications Magazine, 2014, 52(2): 186-195.
[6] MOLISCH A F, RATNAM V V, HAN S, et al. Hybrid Beamforming for Massive MIMO: A Survey[J]. IEEE Communications Magazine, 2017, 55(9): 134-141.
[7] SUN S, LI R, HAN C, et al. How to differentiate between near field and far field: Revisiting the Rayleigh distance[J]. to appear in IEEE Communications Magazine. [Online]. Available: http://export.arxiv.org/abs/2309.13238v2.
[8] RUSEK F, PERSSON D, LAU B K, et al. Scaling Up MIMO: Opportunities and Challenges with Very Large Arrays[J]. IEEE Signal Processing Magazine, 2013, 30(1): 40-60.
[9] HEATH R W, GONZÁLEZ-PRELCIC N, RANGAN S, et al. An Overview of Signal Processing Techniques for Millimeter Wave MIMO Systems[J]. IEEE Journal of Selected Topics in Signal Processing, 2016, 10(3): 436-453.
[10] RANGAN S, RAPPAPORT T S, ERKIP E. Millimeter-Wave Cellular Wireless Networks: Potentials and Challenges[J]. Proceedings of the IEEE, 2014, 102(3): 366-385.
[11] 信科，金思年，陈艺灵，岳殿武，鞠默然. 智能反射面辅助的毫米波大规模 MIMO 技术综述[J]. 移动通信, 2022, 46(12): 12-18.
[12] ZHANG J, YU X, LETAIEF K B. Hybrid Beamforming for 5G and Beyond Millimeter-Wave Systems: A Holistic View[J]. IEEE Open Journal of the Communications Society, 2020, 1: 77-91.
[13] YANG J, ZHU W, SUN S, et al. Deep Learning for Joint Design of Pilot, Channel Feedback, and Hybrid Beamforming in FDD Massive MIMO-OFDM Systems[J]. IEEE Communications Letters, 2024, 28(2): 313-317.
[14] AYACH O E, RAJAGOPAL S, ABU-SURRA S, et al. Spatially Sparse Precoding in Millimeter Wave MIMO Systems[J]. IEEE Transactions on Wireless Communications, 2014, 13(3): 1499-1513.
[15] YU X, SHEN J C, ZHANG J, et al. Alternating Minimization Algorithms for Hybrid Precoding in Millimeter Wave MIMO Systems[J]. IEEE Journal of Selected Topics in Signal Processing, 2016,10(3): 485-500.
[16] GAO X, DAI L, HAN S, et al. Energy-Efficient Hybrid Analog and Digital Precoding for MmWave MIMO Systems With Large Antenna Arrays[J]. IEEE Journal on Selected Areas in Communications, 2016, 34(4): 998-1009.
[17] PARK S, ALKHATEEB A, HEATH R W. Dynamic Subarrays for Hybrid Precoding in Wideband mmWave MIMO Systems[J]. IEEE Transactions on Wireless Communications, 2017, 16(5): 2907-2920.
[18] GADIEL G M, NGUYEN N T, LEE K. Dynamic Unequally Sub-Connected Hybrid Beamforming Architecture for Massive MIMO Systems[J]. IEEE Transactions on Vehicular Technology, 2021,70(4): 3469-3478.
[19] WALDEN R. Analog-to-digital converter technology comparison[C]//Proceedings of 1994

IEEE GaAs IC Symposium. 1994: 217-219.
[20] ORHAN O, ERKIP E, RANGAN S. Low power analog-to-digital conversion in millimeter wave systems: Impact of resolution and bandwidth on performance[C]//2015 Information Theory and Applications Workshop (ITA). 2015: 191-198.
[21] MENENDEZ-RIAL R, RUSU C, GONZALEZ-PRELCIC N, et al. Hybrid MIMO Architectures for Millimeter Wave Communications: Phase Shifters or Switches?[J]. IEEE Access, 2016, 4: 247-267.
[22] GOLDSMITH A, JAFAR S, JINDAL N, et al. Capacity limits of MIMO channels[J]. IEEE Journal on Selected Areas in Communications, 2003, 21(5): 684-702.
[23] 蒲旭敏, 刘雁翔, 孙致南, 李静洁, 陈前斌, 金石. 可重构智能表面中的低复杂度毫米波信道追踪算法[J]. 电子与信息学报, 2023, 45(8): 2911-2918.
[24] RAGHAVAN V, SAYEED A M. Sublinear Capacity Scaling Laws for Sparse MIMO Channels[J]. IEEE Transactions on Information Theory, 2011, 57(1): 345-364.
[25] MARZETTA T L, LARSSON E G, YANG H, et al. Fundamentals of Massive MIMO[M]. Cambridge University Press, 2016.
[26] DUTTA S, BARATI C N, RAMIREZ D, et al. A Case for Digital Beamforming at mmWave[J]. IEEE Transactions on Wireless Communications, 2020, 19(2): 756-770.
[27] ZAPPONE A, JORSWIECK E. Energy Efficiency in Wireless Networks via Fractional Programming Theory[M]. 2015.
[28] YAN L, HAN C, YANG N, et al. Dynamic-Subarray With Fixed Phase Shifters for Energy-Efficient Terahertz Hybrid Beamforming Under Partial CSI[J]. IEEE Transactions on Wireless Communications, 2023, 22(5): 3231-3245.
[29] SUN S, MACCARTNEY G R, RAPPAPORT T S. A novel millimeter-wave channel simulator and applications for 5G wireless communications[C]//2017 IEEE International Conference on Communications (ICC). 2017: 1-7.
[30] SUN S, RAPPAPORT TS, SHAFI M, et al. Propagation Models and Performance Evaluation for 5G Millimeter-Wave Bands[J]. IEEE Transactions on Vehicular Technology, 2018, 67(9): 8422-8439.
[31] MÉNDEZ-RIAL R, RUSU C, GONZÁLEZ-PRELCIC N, et al. Hybrid MIMO Architectures for Millimeter Wave Communications: Phase Shifters or Switches?[J]. IEEE Access, 2016, 4: 247-267.